\documentclass[twocolumn, aps, prb]{revtex4}
\usepackage{amsmath}
\usepackage{amssymb}
\usepackage{graphicx}
\usepackage{footnote}
\usepackage{bm}
\usepackage{dcolumn}
\usepackage[caption=false]{subfig}

\begin{document}

\title{Magnetic oscillations of in-plane conductivity in
quasi-two-dimensional metals}
\author{T. I. Mogilyuk}
\affiliation{National Research Centre ``Kurchatov Institute'', Moscow, 123182, Russia}
\author{P. D. Grigoriev}
\email[Corresponding author; e-mail: ]{grigorev@itp.ac.ru}
\affiliation{L. D. Landau Institute for Theoretical Physics, Chernogolovka, 142432, Russia}
\affiliation{National University of Science and Technology ``MISiS'', Moscow, 119049, Russia}
\date{\today}

\begin{abstract}
We develop the theory of transverse magnetoresistance in layered
quasi-two-dimensional metals. Using the Kubo formula and harmonic expansion,
we calculate intralayer conductivity in a magnetic field perpendicular to
conducting layers. The analytical expressions for the amplitudes and phases
of magnetic quantum oscillations (MQO) and of the so-called slow
oscillations (SlO) are derived and applied to analyze their behavior as a
function of several parameters: magnetic field strength, interlayer transfer
integral and the Landau-level width. Both the MQO and SlO of intralayer and
interlayer conductivities have approximately opposite phase in weak magnetic
field and the same phase in strong field. The amplitude of SlO of intralayer
conductivity changes sign at $\omega_c\tau_0=\sqrt{3}$. There are several
other qualitative difference between magnetic oscillations of in-plane and
out-of-plane conductivity. The results obtained are useful to analyze
experimental data on magnetoresistance oscillations in various strongly
anisotropic quasi-2D metals.
\end{abstract}

\maketitle

\section{Introduction}

Magnetic quantum oscillations (MQO) is a powerful tool for studying
electronic dispersion and Fermi surface geometry of metallic compounds\cite%
{Abrik,Ziman,Shoenberg}. Last decades it is actively used to investigate the
electronic structure of strongly anisotropic layered compounds, including
organic metals (see, e.g., Refs. [\onlinecite{MarkReview2004,Singleton2000Review,OMRev,KartPeschReview,MQORev,LebedBook}]
for reviews), high-temperature superconductors\cite%
{HusseyNature2003,ProustNature2007,SebastianNature2008,AudouardPRL2009,SingletonPRL2010,SebastianPNAS2010,SebastianPRB2010,SebastianPRL2012,SebastianNature2014,ProustNatureComm2015}
(reviewed in Refs. [\onlinecite{SebastianRepProgPhys2012,ProustComptesRendus2013,AnnuReviewYBCO2015}]), etc.
In layered compounds magnetoresistance has several new and useful
qualitative effects, which do not appear in almost isotropic 3D metals. The
theory of magnetoresistance in 2D metals\cite{AndoFowlerSternRMP82,QHERMP1995}, extensively developed in connection to
quantum Hall effect, is also inapplicable to quasi-2D  (Q2D) metals even a
weak interlayer hopping changes drastically the 2D localization effects and
most electronic properties.

The Fermi surface (FS) of layered metals, e.g., corresponding to the
electron dispersion in Eq. (\ref{eq:e3D0}), is a warped cylinder. Such a FS
has two close extremal cross-section areas $S_{1}$ and $S_{2}$ by the planes
in $k$-space perpendicular to magnetic field $\boldsymbol{B}$, which give
two close MQO frequencies $F_{1,2}=S_{1,2}/(2\pi e\hbar )$. According to the
standard theory\cite{Abrik,Ziman,Shoenberg}, the observed MQO are given by
the sum of oscillations with these two frequencies and almost equal
amplitudes, which gives the beats of MQO amplitude\cite{Shoenberg}, typical
to Q2D metals. The beat frequency 
\begin{equation}
\Delta F\equiv F_{1}-F_{2}\approx 2t_{z}B_{z}/(\hbar \omega _{c}),
\label{eq:Fb}
\end{equation}
can be used to measure the interlayer transfer integral $t_{z}\approx \Delta
F\hbar \omega _{c}/(2B_{z})$, while its nontrivial dependence on the tilt
angle $\theta $ of magnetic field (with respect to the normal to conducting
layers), given by\cite{Yam} 
\begin{equation}
\Delta F\left( \theta \right) /\Delta F\left( 0\right) =J_{0}\left(
k_{F}d\tan \theta \right) ,  \label{eq:BesselAMRO}
\end{equation}
allows to extract the in-plane Fermi momentum $k_{F}$. As follows from Eq. (%
\ref{eq:BesselAMRO}), the beat frequency $\Delta F\left( \theta \right) $ goes
to zero in the so-called Yamaji angles $\theta _{Yam}$, given by the zeros
of the Bessel function: $J_{0}\left( k_{F}d\tan \theta _{Yam}\right) =0$.
The angular oscillations of the effective interlayer transfer integral $%
t_{z}\left( \theta \right) $, given by Eq. (\ref{eq:BesselAMRO}), also result
in the angular magnetoresistance oscillations (AMRO), first discovered\cite%
{KartsAMRO1988} in Q2D organic metal $\beta $-(BEDT-TTF)$_{2}$IBr$_{2}$ in
1988 and then actively studied both in Q2D and Q1D organic metals\cite{MarkReview2004,Singleton2000Review,OMRev,KartPeschReview,MQORev,LebedBook,Yagi1990,Kurihara,MosesMcKenzie1999,TarasPRB2014}.
The interplay between AMRO and MQO is also nontrivial\cite%
{TarasPRB2014,TarasPRB2017} and leads to some new effects, such as ``false
spin zeros''\cite{TarasPRB2017}.

Another interesting feature of magnetoresistance in Q2D metals is the so-called
slow oscillations (SlO)\cite{SO,Shub}. These oscillations come from the
mixing of two close frequencies $F_{1}$ and $F_{2}$ and have the frequency
equal to the doubled beat frequency in Eq. (\ref{eq:Fb}). Similarly to AMRO and
contrary to the usual MQO, the SlO are not sensitive to the smearing of the
Fermi level, because they contain only the difference of Fermi levels at
different $k_{z}$ given by $t_{z}$. Hence, the SlO are usually much stronger
than the true MQO and can be observed at much higher temperature\cite%
{SO,GrigEuroPhys2016}. These slow oscillations were first observed in layered
organic metal $\beta $-(BEDT-TTF)$_{2}$IBr$_{2}$ and erroneously interpreted
as MQO from small FS pockets\cite{Kar1,Kar2}. Similar oscillations have also
been observed in other organic conductors, e.g., $\beta $-(BEDT-TTF)$_{2}$I$%
_{3}$ \cite{Kar3,Wosn2}, $\kappa $-(BEDT-TTF)$_{2}$Cu$_{2}$(CN)$_{3}$ \cite{Ohmi}, and $\kappa $-(BEDT-TSF)$_{2}$C(CN)$_{3}$ \cite{togo}, while the
band structure calculations\cite{OMRev} do not predict the corresponding
small FS pockets in these compounds. The $k_{z}$ dispersion is not the only
possible source of SlO. In fact, any splitting of the electron dispersion,
leading to two close FS extremal cross-section areas, produces slow
oscillations of MR with frequency given by the double difference between
these FS areas. For example, the bilayer crystal structure, common in many
cuprate high-Tc superconductors and in numerous other strongly anisotropic
materials, produces such splitting of electron spectrum and the
corresponding SlO\cite{GrigEuroPhys2016,MQOYBCOJETPL,MQOYBCOPRB}.

The SlO turn out to be quite useful to study the parameters of electronic
structure of layered metals. First, their frequency $F_{SlO}=2\Delta F$
gives the difference between the two close extremal FS cross-section areas.
Depending on the origin of SlO, this gives the strength of FS warping due to 
$k_{z}$ dispersion and the value of the interlayer transfer integral $t_{z}$
according to Eq. (\ref{eq:Fb}), the bilayer splitting or another type of
splitting of electron spectrum. Second, the Dingle temperature $T_{D}^{\ast}$
of SlO is considerably less than the Dingle temperature $T_{D}$ of MQO\cite{SO}, because at low temperature it only contains the contribution from
short-range impurities and does not contain the variations of the Fermi
level due to long-range spatial inhomogeneities that damp MQO. Hence, the
comparison of the Dingle temperatures of SlO and MQO gives information about
the type of disorder. In typical samples of organic metal $\beta$-(BEDT-TTF)$%
_{2}$I$_{3}$ the ratio $T_{D}/T_{D}^{\ast}\approx5.3\gg1$\cite{SO}, which
makes SlO much stronger than MQO at any temperature. Third, if SlO are due
to $k_{z}$ dispersion, the angular dependence of SlO frequency gives the
in-plane Fermi momentum $k_{F}$ according to Eq. (\ref{eq:BesselAMRO}).

In addition to SlO, in Q2D metals there is another notable effect of a phase
shift of the beats of MQO of interlayer conductivity as compared to magnetization\cite{PhSh}. This phase shift increases with the increase of magnetic field. The
explanation and calculation of this effect\cite{PhSh,Shub}, done using the
Boltzmann transport equation and the Kubo formula, has shown that, similarly
to SlO, it appears when the terms $\sim\hbar\omega_{c}/t_{z}$ are not
neglected. Hence, in almost isotropic 3D metals, where $t_{z}$ is of the
order of Fermi energy $E_{F}\gg\hbar\omega_{c}$, both effect are negligibly
small. However, in Q2D conductors, where $\hbar\omega_{c}/t_{z}\sim1$, both
effects can be strong.

The rigorous theory of SlO was developed only for the interlayer
magnetoresistance\cite{SO,Shub}. However, their quite generic origin and
various experiments\cite%
{SebastianNature2008,ProustNatureComm2015,GrigEuroPhys2016} suggest that
similar SlO must also be observed in the in-plane electronic transport. A
semi-phenomenological description of in-plane SlO, proposed in Refs. [\onlinecite{GrigEuroPhys2016,MQOYBCOPRB}], does not contain the calculation of in-plane
diffusion coefficient $D_{||}$ but only assumes that its oscillations have
the same phase as the oscillations of the density of states (DoS) due to the
Landau quantization. Even this is not generally valid, as we show below. In
addition, in Refs. [\onlinecite{GrigEuroPhys2016,MQOYBCOPRB,Korsh}] the amplitude of MQO
of $D_{||}$, which affects the amplitude and even the sign of SlO of
intralayer MR, has not been calculated.

In this paper we calculate the in-plane MR in layered Q2D metals using the
Feynman diagram technique. This calculation shows some qualitative
differences of intralayer and interlayer MR. For example, the amplitude of
SlO turns out to have non-monotonic magnetic-field dependence and may even
change the sign. The phase shifts of MQO and their beats in Q2D metals also
differ for intralayer and interlayer MR.

\section{The model and basic formulas}

Let's consider layered Q2D metals with electron dispersion 
\begin{equation}
\epsilon_{3D}(\boldsymbol{k})=\hbar^2\left(k_{x}^{2}+k_{y}^{2}\right)/(2m_{%
\ast})-2t_{z}\cos(k_{z}d),  \label{eq:e3D0}
\end{equation}
where the interlayer transfer integral $t_{z}$\ is assumed to be independent
of electron momentum\footnote{%
The case $t_{z}\left(\boldsymbol{k}_{\parallel}\right)\neq const$ was also
studied\cite{Bergemann,GrigAMRO2010,Mark92}.}. In a magnetic
field $\boldsymbol{B}$ along the $z$-axis, i.e., perpendicular to conducting
layers, its electron dispersion becomes 
\begin{equation}
\epsilon(n,\,k_{z})=\hbar\omega_{c}\left(n+\frac{1}{2}\right)-2t_{z}%
\cos(k_{z}d),  \label{eq:e3D}
\end{equation}
where $\omega_{c}=eB_{z}/(m_{\ast}c)$ is cyclotron frequency, $m_{\ast}$ is
effective electron mass, $e$ is the electric charge, and $c$ is the speed of
light. The diagonal component of the in-plane conductivity tensor $%
\sigma_{ij}(\varepsilon)$ is given by \cite%
{Economou,AGD,Kurihara,Streda,Bastin} 
\begin{eqnarray}
\sigma_{xx}(\varepsilon) & = & \frac{e^{2}\hbar}{\pi V}\sum_{\left\{
n,\,n^{\prime}\right\} =0}^{+\infty}\sum_{k_{x},\,k_{z}}|\left\langle
n^{\prime},\,k_{x},\,k_{z}|v_{x}|k_{z},\,k_{x},\,n\right\rangle |^{2}\times 
\notag \\
& & \times\text{Im}G_{n^{\prime}}^{R}(k_{x},\,k_{z},\,\varepsilon)\text{Im}%
G_{n}^{R}(k_{x},\,k_{z},\,\varepsilon),  \label{eq:Chiang Kai-shek}
\end{eqnarray}
where $V=L_{x}L_{y}L_{z}$ is the volume, which is cancelled after the
summation over momenta, $\text{Im}G_{n}^{R}$ represents the imaginary part
of the retarded electron Green's function $G_{n}^{R}$, $v_{x}$ is the
electron velocity along $x$ axis. The matrix elements $\left\langle
n^{\prime},\,k_{x},\,k_{z}|v_{x}|k_{z},\,k_{x},\,n\right\rangle $ of
electron velocity $v_{x}=p_{x}/m_{\ast}$ in the basis of the Landau-gauge
quantum numbers $\left\{ k_{x},\,k_{z},\,n\right\} $ of an electron in
magnetic field are given by \cite{Kurihara}

\begin{gather}
\left\langle n^{\prime},\,k_{x},\,k_{z}|v_{x}|k_{z},\,k_{x},\,n\right\rangle
=  \notag \\
=\frac{-i\hbar}{\sqrt{2}m_{\ast}l_{H}}(\sqrt{n^{\prime}+1}%
\delta_{n,n^{\prime}+1}-\sqrt{n^{\prime}}\delta_{n,n^{\prime}-1}),
\label{eq:Faucille}
\end{gather}
where $l_{H}=\sqrt{\hbar c/(eB_{z})}=\sqrt{\hbar/(m_{\ast}\omega_{c})}$ is
the magnetic length. Eq. (\ref{eq:Faucille}) can be checked by a direct
calculation. The square of this matrix element of electron velocity is 
\begin{equation}
|\left\langle n-1,\,k_{x},\,k_{z}|v_{x}|k_{z},\,k_{x},\,n\right\rangle |^{2}=%
\frac{\hbar^{2}n}{2m_{\ast}^{2}l_{H}^{2}}.  \label{eq:v2}
\end{equation}
The summation over momenta in Eq. (\ref{eq:Chiang Kai-shek}) can be replaced by
the integration according to: 
\begin{equation}
\sum_{k_{x}}=\int_{0}^{L_{y}/l_{H}^{2}}\frac{dk_{x}L_{x}}{2\pi}%
,~\sum_{k_{z}}=\int_{-\pi/d}^{\pi/d}\frac{dk_{z}L_{z}}{2\pi}.
\label{eq:Castro}
\end{equation}

In the Born approximation or even in the self-consistent Born approximation
(SCBA) the self-energy part $\Sigma ^{R}(\varepsilon )$ from short-range
impurity scattering depends only on electron energy $\varepsilon $ and does
not depend on electron quantum numbers\cite{Shub,ChampelMineev,WIPRB2011,GrigPRB2013}$^,$ \footnote{This property for the scattering by point-like impurities can be proved even
in the ``non-crossing'' approximation.\cite{WIPRB2011}},
 and the
electron Green's function does not depend on $k_{x}$: 
\begin{gather}
\text{Im}G_{n}^{R}(k_{x},\,k_{z},\,\varepsilon )=\text{Im}%
G_{n}^{R}(k_{z},\,\varepsilon )=  \notag \\
=\frac{\text{Im}\Sigma ^{R}(\varepsilon )}{\left[ \varepsilon-\epsilon
_{n}+2t_{z}\cos (k_{z}d)-\text{Re}\Sigma ^{R}(\varepsilon )\right] ^{2}+%
\left[ \text{Im}\Sigma ^{R}(\varepsilon )\right] ^{2}},
\label{eq:Mao Zedong}
\end{gather}%
where $\epsilon _{n}=\hbar \omega _{c}(n+1/2)$. Substituting Eqs. (\ref{eq:v2}--%
\ref{eq:Mao Zedong}) to Eq. (\ref{eq:Chiang Kai-shek}) one obtains the
expression for diagonal conductivity in the SCBA approximation: 
\begin{gather}
\sigma _{xx}(\varepsilon )=\frac{e^{2}(\hbar \omega _{c})^{2}\Gamma ^{2}}{%
4\pi ^{3}\hbar }\int_{-\pi /d}^{\pi /d}dk_{z}\times  \notag \\
\times \sum_{n=0}^{+\infty }\frac{n\left[ (\epsilon _{n+1}-\varepsilon
_{\ast }-2t_{z}\cos (k_{z}d))^{2}+\Gamma ^{2}\right] ^{-1}}{\left[
\epsilon _{n}-\varepsilon _{\ast }-2t_{z}\cos (k_{z}d)\right] ^{2}+\Gamma
^{2} },  \label{eq:Sun-Yixian}
\end{gather}%
where we introduced the notations: 
\begin{equation}
\varepsilon _{\ast }\equiv \varepsilon -\text{Re}\Sigma ^{R}(\varepsilon
),~~\Gamma \equiv \left\vert \text{Im}\Sigma ^{R}(\varepsilon )\right\vert .
\label{eq:eps}
\end{equation}%
Introducing the dimensionless quantities 
\begin{gather}
\alpha\equiv\alpha (\varepsilon _{\ast })\equiv 2\pi \varepsilon _{\ast
}/(\hbar \omega _{c}),\, a\equiv \alpha (\varepsilon
_{\ast })+\lambda \cos (k_{z}d),
\label{eq:alpha_gamma} \\
\lambda =4\pi t_{z}/(\hbar \omega _{c}),\,\gamma =2\pi \Gamma /(\hbar \omega _{c}),  \label{eq:lambda_a}
\end{gather}
we can rewrite the expression (\ref{eq:Sun-Yixian}) for diagonal
conductivity as

\begin{equation}
\sigma_{xx}(\varepsilon)=\frac{e^{2}\gamma^{2}}{\pi\hbar}\int_{-\pi/d}^{%
\pi/d}dk_{z}\sum_{n=0}^{+\infty}f\left(n\right),  \label{eq:sxxSLL}
\end{equation}
where 
\begin{equation}
f\left(n\right)\equiv\frac{n\left(\left[2\pi\left(n-\frac{1}{2}\right)-a\right]%
^{2}+\gamma^{2}\right)^{-1}}{\left[2\pi\left(n+\frac{1}{2}\right)-a\right]%
^{2}+\gamma^{2}}.  \label{eq:fn}
\end{equation}

\section{Harmonic expansion of conductivity}

The sum over the LL index $n$ in Eq. (\ref{eq:sxxSLL}) can be transformed to
the sum over harmonics using the Poisson summation formula\cite{Wilton},
given by 
\begin{equation}
\sum_{n=0}^{+\infty}f(n)=\sum_{p=-\infty}^{+\infty}\int_{h}^{+\infty}dnf(n)%
\exp(2\pi ipn),  \label{eq:Poisson}
\end{equation}
where the number $h\in\left(-1,\,0\right)$. In the limit of strong harmonic
damping, i.e., when the factor $R_{D}J_{0}\left(\lambda\right)\ll1$, where $%
R_{D}=\exp(-\gamma)\approx
R_{D0}=\exp(-\gamma_{0})=\exp(-2\pi^{2}k_{B}T_{D}/(\hbar\omega_{c}))$ is the
Dingle factor, we may keep only the zeroth and first harmonics in this
expansion: 
\begin{equation}
\sigma_{xx}(\varepsilon)\approx\sigma_{xx}^{(0)}(\varepsilon)+%
\sigma_{xx}^{(1)}(\varepsilon),  \label{eq:sxxP}
\end{equation}
where the zero-harmonic term 
\begin{equation}
\sigma_{xx}^{(0)}(\varepsilon)=\frac{e^{2}\gamma^{2}}{\pi\hbar}%
\int_{-\pi/d}^{\pi/d}dk_{z}\int_{-1/2}^{+\infty}dnf\left(n\right),
\label{eq:s0}
\end{equation}
and the first-harmonic term 
\begin{equation}
\sigma_{xx}^{(1)}(\varepsilon)=2\frac{e^{2}\gamma^{2}}{\pi\hbar}%
\int_{-\pi/d}^{\pi/d}dk_{z}\int_{-1/2}^{+\infty}dnf\left(n\right)\cos\left(2%
\pi n\right).  \label{eq:s1}
\end{equation}

The integrals in Eqs. (\ref{eq:s0}) and (\ref{eq:s1}) simplify in the limit
when the number $n_{F}$ of filled LLs is large, i.e., when $a\sim
E_{F}/(\hbar \omega _{c})\approx n_{F}\gg 1$, where $E_{F}$ is the Fermi
energy. Then, after changing the integration variable from $n$ to $l=2\pi
\left( n+1/2\right) -a$, we can also change the lower integration limit from 
$-a$ to $-\infty $, because all integrals converge at lower integration
limit. The integral over $n$ in Eq. (\ref{eq:s0}) becomes 
\begin{eqnarray}
\int_{-1/2}^{+\infty }dnf\left( n\right) &\approx &\int_{-\infty }^{+\infty }%
\frac{dl(l+a-\pi )/\left( 2\pi \right) ^{2}}{(l^{2}+\gamma
^{2})((l-1)^{2}+\gamma ^{2})}=  \notag \\
&=&\frac{a}{8\pi \gamma (\gamma ^{2}+\pi ^{2})},  \label{eq:I0m}
\end{eqnarray}%
and substituting this to Eq. (\ref{eq:s0}) we obtain 
\begin{eqnarray}
\sigma _{xx}^{(0)}(\varepsilon ) &=&\frac{e^{2}\gamma }{8\pi ^{2}\hbar }%
\int_{-\pi /d}^{\pi /d}dk_{z}\frac{\lambda \cos (k_{z}d)+\alpha }{\gamma
^{2}+\pi ^{2}}=  \notag \\
&=&\frac{e^{2}}{4\pi \hbar d}\frac{\alpha \gamma }{\gamma ^{2}+\pi ^{2}}.
\label{eq:sxxP0}
\end{eqnarray}%
Similarly, at $a\gg 1$ the integration over $n$ in expression (\ref{eq:s1})
for $\sigma _{xx}^{(1)}(\varepsilon )$ gives 
\begin{equation}
\int_{-1/2}^{+\infty }dnf\left( n\right) \cos \left( 2\pi n\right) \approx 
\frac{a\gamma \cos (a)}{8\pi (\gamma ^{2}+\pi ^{2})}\exp (-\gamma ).
\label{eq:I1m}
\end{equation}%
Substituting this and Eq. (\ref{eq:alpha_gamma}) to Eq. (\ref{eq:s1}), we obtain the integral over $k_{z}$ only,
which can be easily taken: 
\begin{gather}
\sigma _{xx}^{(1)}(\varepsilon )=-\frac{e^{2}\gamma }{4\pi ^{2}\hbar }%
\int_{-\pi /d}^{\pi /d}dk_{z}\frac{\alpha (\varepsilon _{\ast })+\lambda
\cos (k_{z}d)}{\gamma ^{2}+\pi ^{2}}\times  \notag \\
\times \cos \left[ \alpha (\varepsilon _{\ast })+\lambda \cos (k_{z}d)\right]
\exp (-\gamma )=  \notag \\
=-\frac{e^{2}\alpha }{2\pi \hbar d}\frac{\gamma \exp (-\gamma )}{\gamma
^{2}+\pi ^{2}}\left[ J_{0}(\lambda )\cos \left( \alpha \right) -\frac{%
\lambda }{\alpha }J_{1}(\lambda )\sin (\alpha )\right] ,  \label{eq:s1b}
\end{gather}%
where to integrate over $k_{z}$ we used the identities\cite{Grad,Prudnikov}: 
\begin{eqnarray}
\int_{-\pi }^{\pi }dn\exp (ia\cos (n)) &=&2\pi J_{0}(a), \\
\int_{-\pi }^{\pi }dn\cos (n)\exp (ia\cos (n)) &=&2\pi iJ_{1}(a).
\end{eqnarray}%
If $\lambda /\alpha \approx 2t_{z}/E_{F}\ll 1$, in Eq. (\ref{eq:s1b}) one can neglect the last
term in the square brackets, but at $2t_{z}/E_{F}\sim 1$ it must be kept.
This term gives the phase shift of MQO of conductivity and leads to the
finite amplitude of MQO even in the beat nodes (see Eq. (\ref{eq:rakshasa})
below), which can be used to measure the ratio $2t_{z}/E_{F}$.

In the SCBA for point-like impurity scattering the electron self-energy is
proportional to the Green's function in the coinciding points $%
G(r,\,r,\,\varepsilon)$, and its oscillations are given by\cite{Shub}

\begin{equation}
\frac{\Sigma ^{R}(\varepsilon )}{\Gamma _{0}}=A(\varepsilon
)-i-2i\sum_{p=1}^{+\infty }(-1)^{p}\exp \left[ p(i\alpha (\varepsilon _{\ast
})-\gamma )\right] J_{0}\left( \lambda p\right) ,  \label{eq:S1}
\end{equation}%
where $\Gamma _{0}$ is a non-oscillating part of Im$\Sigma ^{R}(\varepsilon
) $, related to mean free time $\tau _{0}=\hbar /(2\Gamma _{0})$ without
magnetic field, and $A(\varepsilon )$ is a slowly-varing function of energy $%
\varepsilon $, which only shifts the chemical potential. Hence, $%
A(\varepsilon )$ does not affect the observed conductivity and is
hereinafter neglected.

Below we find explicitly all the terms which contribute to MQO and SlO in
the lowest order in the small factor $R_{D}J_{0}\left(\lambda\right) $.

\subsection{Contribution from the zero-harmonic term $\protect\sigma%
_{xx}^{(0)}$}

Eq. (\ref{eq:sxxP0}) is an oscillating function of $\varepsilon $, because
it contains oscillating functions $\gamma (\varepsilon )$ and $\alpha
(\varepsilon _{\ast })$. Keeping only zeroth and first harmonics in Eq. (\ref%
{eq:S1}) we obtain 
\begin{equation}
\gamma =\gamma (\varepsilon )\approx \gamma _{0}\left[ 1-2\exp \left(
-\gamma \right) \cos \left( \alpha \right) J_{0}\left( \lambda \right) %
\right] ,  \label{eq:g1}
\end{equation}%
and $\alpha =\alpha (\varepsilon _{\ast })$ also contains oscillations
coming from Re$\Sigma ^{R}(\varepsilon )$ in Eq. (\ref{eq:S1}). However, the
relative amplitude of $\alpha (\varepsilon _{\ast })$ oscillations is much
smaller, namely by a factor $\gamma _{0}/\alpha\approx\Gamma _{0}/E_{F}\ll 1$,
than that of $\gamma (\varepsilon )$, although their absolute amplitudes are
comparable. Hence, in Eq. (\ref{eq:sxxP0}) the oscillations of $\alpha
(\varepsilon _{\ast })$ can be neglected. Note that the products $\cos
\left( \alpha \right) e^{-\gamma }$ and $\sin \left( \alpha \right)
e^{-\gamma }$ do not give the SlO in the second order in $R_{D}$. Indeed,
using Eq. (\ref{eq:S1}) and introducing the small parameter 
\begin{equation}
\gamma _{1}\equiv 2\gamma _{0}R_{D}J_{0}\left( \lambda \right) \ll 1,
\label{eq:gamma1}
\end{equation}%
in the second order in $R_{D}$ we obtain 
\begin{equation}
\cos \left( \alpha \right) \approx \cos \left[ \overline{\alpha}+\gamma _{1}\sin
\left( \overline{\alpha}\right) \right] \approx \cos \left[ \overline{\alpha}\right]
-\gamma _{1}\sin ^{2}\left( \overline{\alpha}\right) ,
\end{equation}%
where $\overline{\alpha}=2\pi \overline{\varepsilon}_{\ast }/(\hbar \omega _{c})=2\pi
E_{F}/(\hbar \omega _{c})$ is the value of $\alpha $ averaged over MQO
period, and 
\begin{equation}
\exp (-\gamma )\approx R_{D}\exp \left[ \gamma _{1}\cos \left( \overline{\alpha}%
\right) \right] \approx R_{D}\left[ 1+\gamma _{1}\cos \left( \overline{\alpha}%
\right) \right] .
\end{equation}%
In the second order in $R_{D}$ the product 
\begin{gather}
\cos \left( \alpha \right) \exp (-\gamma )\approx R_{D}\left( \cos \left[ 
\overline{\alpha}\right] +\gamma _{1}\left[ \cos ^{2}\left( \overline{\alpha}\right)
-\sin ^{2}\left( \overline{\alpha}\right) \right] \right) =  \notag \\
=R_{D}\left( \cos \left[ \overline{\alpha}\right] +\gamma _{1}\cos \left[ 2\overline{%
\alpha}\right] \right)  \label{eq:PSO}
\end{gather}%
does not contain the constant term giving SlO but only the second harmonics $%
\cos \left[ 2\overline{\alpha}\right] $. Similarly, 
\begin{equation}
\sin \left( \alpha \right) \approx \sin \left[ \overline{\alpha}+\gamma _{1}\sin
\left( \overline{\alpha}\right) \right] \approx \sin \left[ \overline{\alpha}\right]
+\gamma _{1}\cos \left( \overline{\alpha}\right) \sin \left( \overline{\alpha}\right)
\end{equation}%
and $\sin \left( \alpha \right) \exp (-\gamma )$ do not contain constant or
SlO terms in the second order in $R_{D}J_{0}\left( \lambda \right) $. Hence,
in the second order in $R_{D}$, Eq. (\ref{eq:g1}) simplifies to 
\begin{equation}
\gamma \approx \gamma _{0}\left[ 1-2\exp \left( -\gamma _{0}\right)
J_{0}\left( \lambda \right) \cos \left( \overline{\alpha}\right) \right] .
\label{eq:g1m}
\end{equation}

Substituting Eq. (\ref{eq:g1m}) to (\ref{eq:sxxP0}), expanding up to the
second order in $R_{D}J_{0}\left(\lambda\right)$ and replacing $\alpha$ with 
$\overline{\alpha}$ we obtain \footnote{In fast quantum oscillations we keep only the first-order term in the Dingle
factor. However, in slow oscillations we keep also the second-order terms,
because contrary to fast quantum oscillations they are not damped by
temperature and sample inhomogeneties.}: 
\begin{equation}
\sigma_{xx}^{(0)}(\varepsilon)=\overline{\sigma}_{xx}^{(0)}(\varepsilon)+%
\sigma_{xx}^{(0)}(\varepsilon)_{qo}+\sigma_{xx}^{(0)}(\varepsilon)_{so},
\label{eq:s0t}
\end{equation}
where the non-oscillating Drude conductivity 
\begin{equation}
\overline{\sigma}_{xx}^{(0)}(\varepsilon)=\overline{\sigma}_{xx}^{(0)}\approx\frac{%
e^{2}}{4\pi\hbar d}\frac{\overline{\alpha}\gamma_{0}}{\gamma_{0}^{2}+\pi^{2}},
\label{eq:sxxDrude}
\end{equation}
the fast quantum oscillations of conductivity come from the first-order term
in $R_{D}J_{0}\left(\lambda\right)$ and are given by 
\begin{equation}
\sigma_{xx}^{(0)}(\varepsilon)_{qo}\approx2\overline{\sigma}_{xx}^{(0)}R_{D}J_{0}%
\left(\lambda\right)\frac{\gamma_{0}^{2}-\pi^{2}}{\gamma_{0}^{2}+\pi^{2}}%
\cos\left(\overline{\alpha}\right),  \label{eq:s0qo}
\end{equation}
and the slow oscillations of conductivity appear in the second order in $%
R_{D}J_{0}\left(\lambda\right)$: 
\begin{equation}
\sigma_{xx}^{(0)}(\varepsilon)_{so}\approx2\overline{\sigma}_{xx}^{(0)}\frac{%
\gamma_{0}^{2}(\gamma_{0}^{2}-3\pi^{2})}{(\gamma_{0}^{2}+\pi^{2})^{2}}
R_{D}^{2}J_{0}^{2}\left(\lambda\right),  \label{eq:s0so}
\end{equation}
where we have used the identity $\cos^{2}\left(\overline{\alpha}\right)=\left[
1+\cos\left(2\overline{\alpha}\right)\right]/2$ and neglected the second harmonics of
MQO, i.e., omitted terms $\propto\cos\left(2\overline\alpha\right)$.

\subsection{Contribution from the first-harmonic term $\protect\sigma%
_{xx}^{(1)}$ and total expressions for magnetic oscillations}

\begin{figure*}[tbh]
\subfloat[\label{fig:Amplitude-of-quantumA} The amplitude of quantum oscillations of normalized in-plane diagonal conductivity $A_{xx}^{qo}d/G_{0}$ given by Eq. (\ref{eq:rakshasa}) as a function of $1/\lambda\sim B_z$ (here $G_{0}=e^{2}/(\pi\hbar)$ is the quantum of conductance) for three different values of Fermi energy $E_F$. The positions of local minima of MQO amplitude (beat nodes) do not depend on Fermi energy $E_F$, but the amplitude of MQO amplitude does according to Eq. (\ref{eq:rakshasa}) and discussion after Eq. (\ref{eq:dphi}). The taken parameters are: $m_{\ast }=0.04m_e$, $\Gamma _{0}=14.5\,K$, $t_{z}=10\, meV$, $E_{F}=\{5,\,10,\, 20\} t_{z}$.]{ \centering{}\includegraphics[width=0.85\textwidth,height=1\columnwidth]{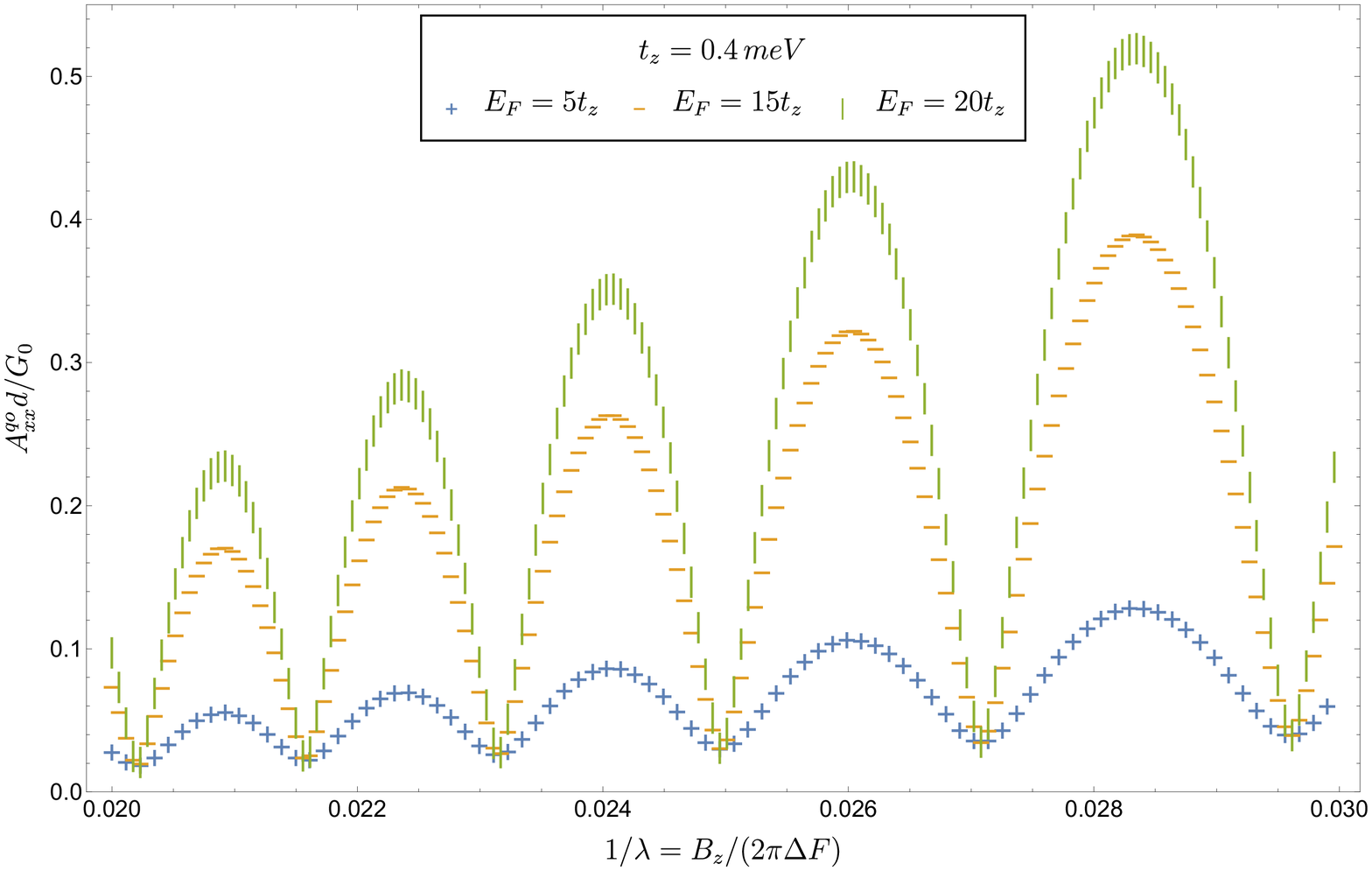}
	} \smallskip{} 
\subfloat[\label{fig:Amplitude-of-quantumB} The same as in Fig. (a) at fixed $E_F$ but for three different values of $t_z$.  The values at local minima of MQO depend on $t_z$. The taken parameters are: $m_{\ast }=0.04m_e$, $\Gamma _{0}=14.5\,K$, $t_{z}=\{1/5,\,1/15,\,1/20\}E_F$, $E_{F}=200\,meV$.]{ \centering{}\includegraphics[width=0.86\textwidth,height=1\columnwidth]{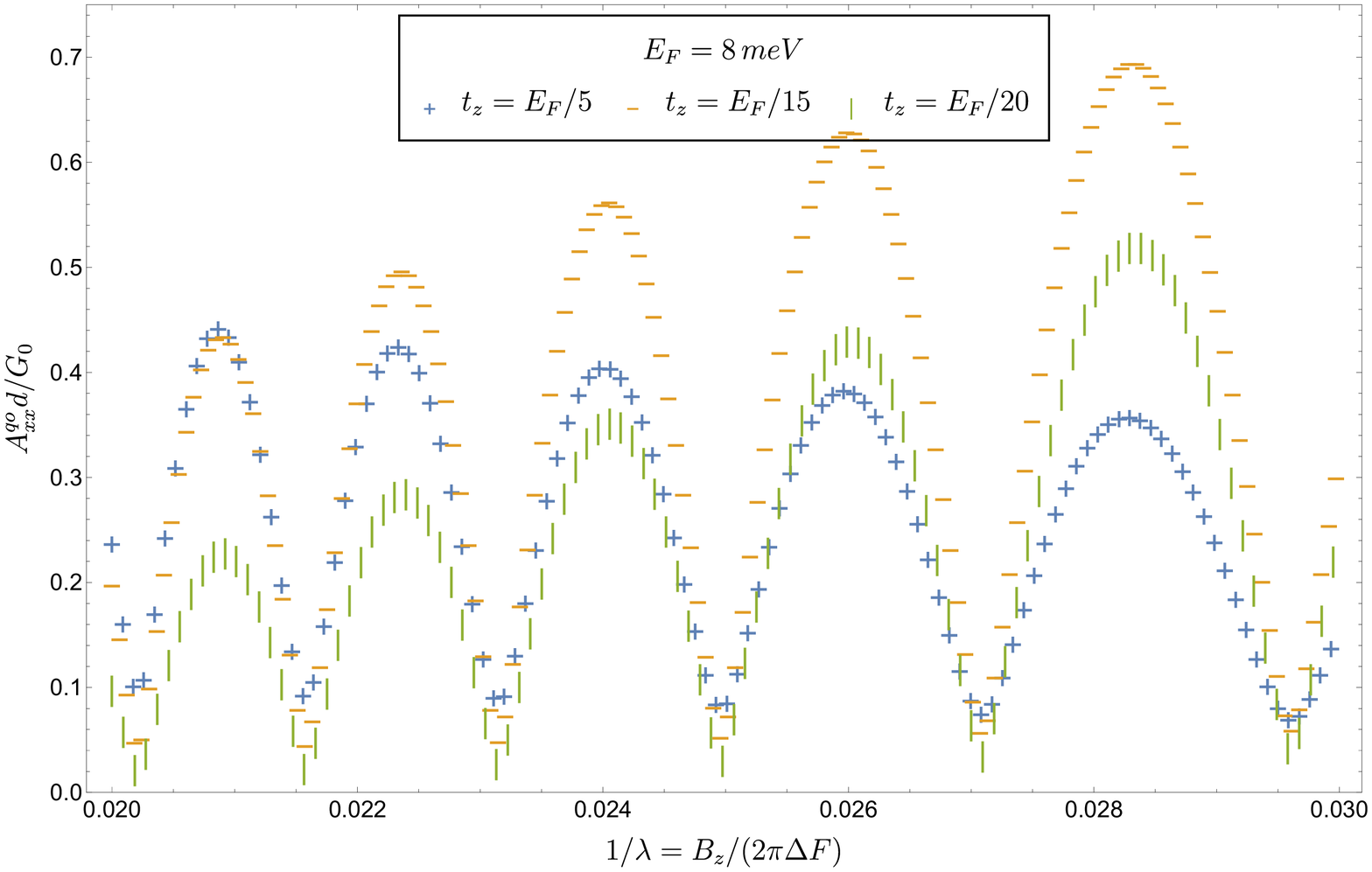}
	}
\caption{The amplitude of quantum oscillations of in-plane diagonal
conductivity for three different ratios $t_z/E_F$.}
\label{fig:Amplitude-of-quantum}
\end{figure*}

To find the fast quantum oscillations of $\sigma _{xx}^{(1)}(\varepsilon )$
in the lowest order in $R_{D}J_{0}(\lambda )$ it is sufficient to replace $%
\gamma $ by $\gamma _{0}$ and $\alpha (\varepsilon _{\ast })$ by its average
value $\overline{\alpha}$ in Eq. (\ref{eq:s1b}): 
\begin{equation}
\sigma _{xx}^{(1)}(\varepsilon )_{qo}\approx-2\overline{\sigma}_{xx}^{(0)}R_{D}%
\left[ J_{0}(\lambda )\cos \left(\overline{\alpha} \right) -\frac{\lambda }{\overline{%
\alpha}}J_{1}(\lambda )\sin (\overline{\alpha} )\right] .  \label{eq:s1b0}
\end{equation}%
Then, the sum of Eqs. (\ref{eq:s0qo}) and (\ref{eq:s1b0}) gives the total
fast quantum oscillations in the first order in $R_{D}J_{0}(\lambda )$: 
\begin{gather}
\sigma _{xx}^{qo}(\varepsilon )=\sigma _{xx}^{(0)}(\varepsilon )_{qo}+\sigma
_{xx}^{(1)}(\varepsilon )_{qo}\approx  \notag \\
\approx-2\overline{\sigma}_{xx}^{(0)}R_{D}\left[ \frac{2\pi ^{2}J_{0}\left(
\lambda \right) }{\gamma _{0}^{2}+\pi ^{2}}\cos \left(\overline{\alpha}\right) -%
\frac{\lambda }{\overline{\alpha}}J_{1}(\lambda )\sin (\overline{\alpha})\right] .
\label{eq:s1qo}
\end{gather}%
We transform this trigonometric expression to 
\begin{equation}
\sigma _{xx}^{qo}(\varepsilon) \approx-A_{xx}^{qo}\cos(\overline{\alpha}+\Delta
\phi _{qo}),
\end{equation}%
where the amplitude of MQO is given by 
\begin{equation}
A_{xx}^{qo}=2\overline{\sigma}_{xx}^{(0)}R_{D}\sqrt{\frac{4\pi ^{4}}{(\gamma
_{0}^{2}+\pi ^{2})^{2}}J_{0}^{2}(\lambda )+\left( \frac{\lambda }{\overline{\alpha%
}}\right) ^{2}J_{1}^{2}(\lambda )}  \label{eq:rakshasa}
\end{equation}
and a phase shift of MQO is 
\begin{gather}
\Delta \phi _{qo}=  \notag \\
=\arccos \left[ \frac{2\pi ^{2}J_{0}\left( \lambda \right) \overline{\alpha}%
}{\sqrt{(2\pi ^{2}J_{0}\left( \lambda \right) \overline{\alpha})^{2}+\left(
\lambda J_{1}(\lambda )\left( \gamma _{0}^{2}+\pi ^{2}\right) \right) ^{2}}}%
\right] .  \label{eq:dphi}
\end{gather}%
This phase shift jumps by $\sim \pi $ and changes the sign of $\sigma
_{xx}^{qo}$ at certain values of magnetic field, corresponding to the beats
of MQO at $J_{0}\left( \lambda \right) =0$. The second term in the
denominator makes this phase jump smoother and is missing in
phenomenological theories\cite{GrigEuroPhys2016,MQOYBCOPRB}. 

\begin{figure*}[tbh]
\centering{}\includegraphics[width=0.86\textwidth,height=1\columnwidth]{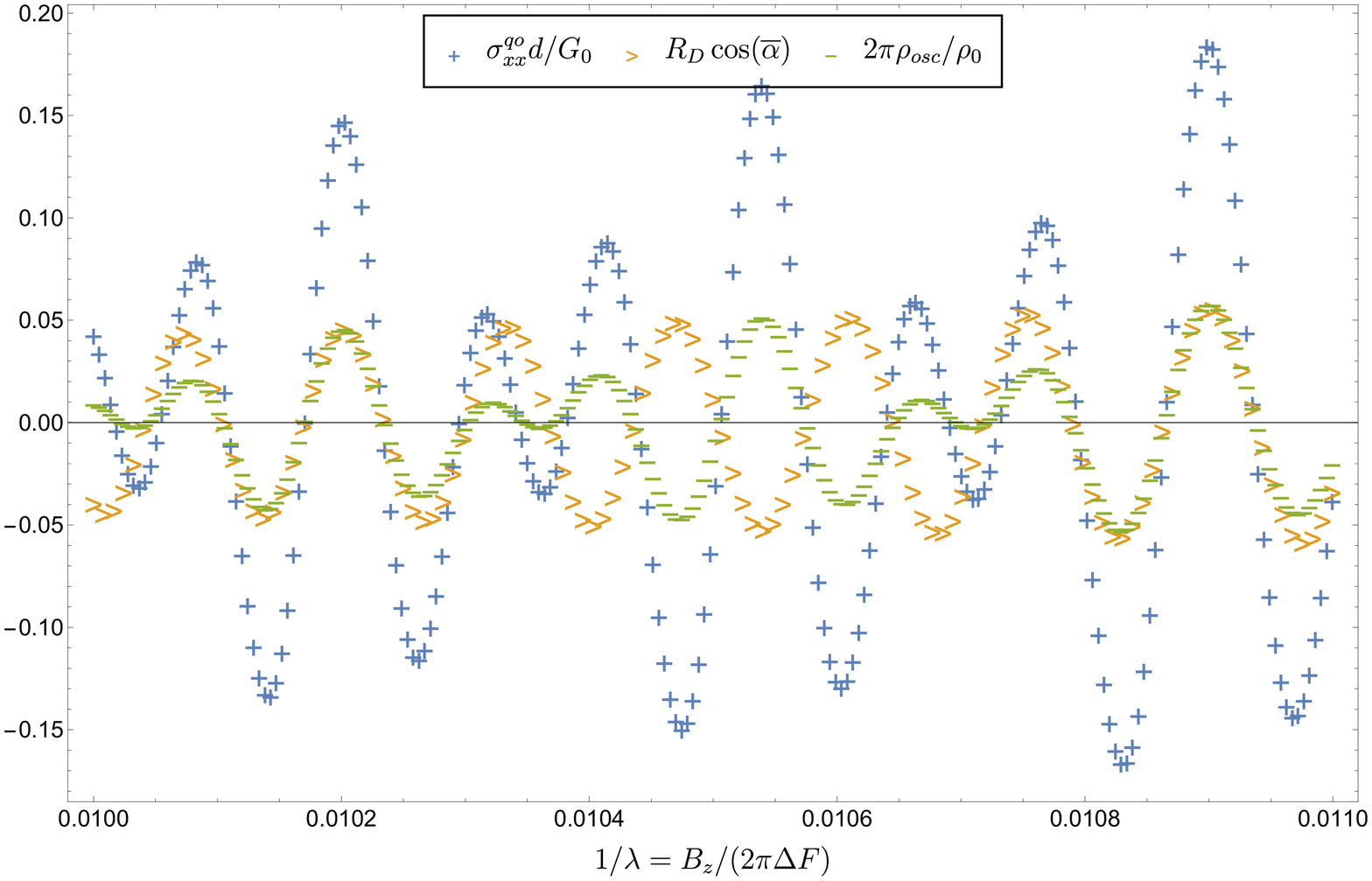}
\caption{Quantum oscillations of in-plane diagonal conductivity as compared
to $R_{D}\cos (\overline\alpha )$ and to the oscillating part $\protect\rho %
_{osc}$ of the DoS, where $\protect\rho _{osc}\equiv -2\protect\rho _{0}\cos
(\overline{\protect\alpha })R_{D}J_{0}(\protect\lambda )$. The phases of DoS and of $\protect\sigma _{xx}$
oscillations coincide everywhere except the proximity of beat nodes. The taken
parameters are: $m_{\ast }=0.04m_{e}$, $\Gamma _{0}=14.5\,K$, $t_{z}=20\,
meV$, $E_{F}=200\,meV$. }
\label{fig:Quantum-oscillations-of-inplane}
\end{figure*}

The derived expressions (\ref{eq:s1qo}--\ref{eq:dphi}), describing the MQO of in-plane
conductivity in the lowest non-vanishing order in $R_{D}J_{0}(\lambda )$,
have several important features. 
Due to the second term in Eq. (\ref{eq:s1qo}), the MQO amplitude $%
A_{xx}^{qo} $, given by Eq. (\ref{eq:rakshasa}) and plotted in Fig. \ref%
{fig:Amplitude-of-quantum}, is nonzero even at beat nodes $J_{0}(\lambda )=0$%
, corresponding to the minima of MQO amplitude, where it increases with the increase of ratio $\lambda /\overline{\alpha}
=2t_{z}/E_{F} $. At maxima the MQO amplitude $%
A_{xx}^{qo}$ is proportional to the square of electron velocity and, for a
parabolic in-plane electron dispersion, to the Fermi energy $E_{F}$, in
agreement with the standard theory\cite{Shoenberg} (see Fig. \ref{fig:Amplitude-of-quantumA}%
). Eqs. (\ref{eq:s1qo}) and (\ref{eq:rakshasa}) suggest that at $\gamma
_{0}\gtrsim \pi $, in addition to the standard Dingle factor, the MQO are
damped by the factor $1/(\gamma _{0}^{2}+\pi ^{2})$. 
We illustrate all this in Fig. \ref{fig:Amplitude-of-quantum}
by plotting the amplitude $A_{xx}^{qo}$ as a function of
 $1/\lambda =B_z/(2\pi\Delta F)$ for three different ratios of $t_{z}/E_{F}$. 
In Fig. \ref{fig:Amplitude-of-quantumA} we keep $t_{z}$ fixed
and vary $E_{F}$, which may correspond to different Fermi-surface pockets or
Fermi-surface reconstruction, and in Fig. \ref{fig:Amplitude-of-quantumB} we
keep $E_{F}$ fixed and vary $t_{z}$. In all figures the
MQO amplitude increases with the increase of magnetic field because of
the Dingle factor. In Fig. \ref{fig:Amplitude-of-quantumA}
at the beat nodes the MQO amplitude is the same for all three curves because 
$E_{F}^{-1}$ in the factor $t_{z}/E_{F}$ is compensated by the overall
factor $E_{F}$ in $\overline{\sigma}_{xx}^{(0)}$. In Fig. \ref{fig:Amplitude-of-quantumB} three different curves, corresponding to various values of $t_{z}$, also
correspond to different magnetic field strength, because we plotted $%
A_{xx}^{qo}$ as function of $1/\lambda \propto B_{z}/t_{z}$. Therefore, at
low field the blue curve, corresponding to $t_{z}=E_{F}/5$, is higher at MQO
maxima. The second term in Eq. (\ref{eq:s1qo}) also results to additional 
phase shift in Eq. (\ref{eq:dphi}), which depends on magnetic field via $\lambda $ and $\gamma _{0}$ and is essential only near the beat nodes $J_{0}\left( \lambda \right) =0$, as illustrated in Fig. \ref{fig:Quantum-oscillations-of-inplane}.
 
To find the slow oscillations of $\sigma_{xx}^{(1)}(\varepsilon)$ in the
lowest (second) order in $R_{D}J_{0}(\lambda)$ we need to expand $\gamma$ in
(\ref{eq:s1b}) according to Eq. (\ref{eq:g1m}) and take into account
oscillations of $\alpha$. This expansion gives 
\begin{gather}
\sigma_{xx}^{(1)}(\varepsilon)-\sigma_{xx}^{(1)}(\varepsilon)_{qo}\approx4%
\overline{\sigma}_{xx}^{(0)}\frac{\pi^{2}-\gamma_{0}^{2}}{%
\gamma_{0}^{2}+\pi^{2}}R_{D}^{2}J_{0}\left(\lambda\right)\times  \notag \\
\times\cos\left(\overline{\alpha}\right)\left[J_{0}(\lambda)\cos\left(\overline{\alpha}%
\right)-\frac{\lambda}{\overline{\alpha}}J_{1}(\lambda)\sin(\overline{\alpha})\right] ,
\end{gather}
and the SlO coming from this expression are given by 
\begin{equation}
\sigma_{xx}^{(1)}(\varepsilon)_{so}\approx2\overline{\sigma}_{xx}^{(0)}(%
\varepsilon)\frac{\pi^{2}-\gamma_{0}^{2}}{\gamma_{0}^{2}+\pi^{2}}%
R_{D}^{2}J_{0}^{2}\left(\lambda\right).  \label{eq:s1SlO}
\end{equation}
The total SlO of diagonal in-plane conductivity are given by the sum of Eqs.
(\ref{eq:s0so}) and (\ref{eq:s1SlO}): 
\begin{equation}
\sigma_{xx}^{so}(\varepsilon)\approx2\pi^{2}\overline{\sigma}%
_{xx}^{(0)}R_{D}^{2}J_{0}^{2}\left(\lambda\right)\frac{\pi^{2}-3%
\gamma_{0}^{2}}{(\gamma_{0}^{2}+\pi^{2})^{2}}.  \label{eq:sSlO}
\end{equation}
To summarize the calculations, the harmonic expansion (by the parameter $%
\gamma_{1}\equiv2\gamma_{0}R_{D}J_{0}\left(\lambda\right)\ll1$) of
intralayer conductivity $\sigma_{xx}(\varepsilon)$ is given by the sum of
three main terms: 
\begin{equation}
\sigma_{xx}(\varepsilon)\approx\overline{\sigma}_{xx}^{(0)}(\varepsilon)+%
\sigma_{xx}^{qo}(\varepsilon)+\sigma_{xx}^{so}(\varepsilon),  \label{eq:Se3}
\end{equation}
where the term $\overline{\sigma}_{xx}^{(0)}(\varepsilon)$ corresponds to the
nonoscillating part of conductivity and is given by Eq. (\ref{eq:sxxDrude}), 
$\sigma_{xx}^{qo}(\varepsilon)$ describes the MQO of intralayer
conductivity, given by Eq. (\ref{eq:s1qo}), and $\sigma_{xx}^{so}(%
\varepsilon)$ describes the slow oscillations of intralayer conductivity and
is given by Eq. (\ref{eq:sSlO}).

\subsection{Damping by temperature and sample inhomogeneities}

As was shown in Refs. [\onlinecite{SO,Shub,MQOYBCOPRB}], the smearing of the Fermi
level by temperature and by long-range sample inhomogeneities damps only the
fast MQO $\sigma_{xx}^{qo}(\varepsilon)$, and does not affect the constant
part $\overline{\sigma}_{xx}^{(0)}(\varepsilon)$ or the SlO $\sigma_{xx}^{so}(%
\varepsilon)$. Indeed, at finite temperature $T$ conductivity $%
\sigma_{xx}=\sigma_{xx}\left(T\right)$ is given by the integral of $%
\sigma_{xx}(\varepsilon)$ over electron energy $\varepsilon$ weighted by the
derivative of Fermi distribution function $n_{F}^{\prime}(%
\varepsilon)=-1/\{4T\cosh^{2}\left[(\varepsilon-\mu)/(2T)\right]\}$ with the chemical potential $\mu =E_F$: 
\begin{equation}
\sigma_{xx}\left(\mu,\,T\right)=\int d\varepsilon\,\left[-n_{F}^{\prime}(%
\varepsilon)\right]\,\sigma(\varepsilon).  \label{eq:sigmaT}
\end{equation}
Among the three terms in Eq. (\ref{eq:Se3}) only the second term $%
\sigma_{xx}^{qo}(\varepsilon)$, describing MQO, is a rapidly oscillating
function of electron energy $\varepsilon$ because of its dependence on $\overline{%
\alpha}\left(\varepsilon\right)$. As a result\ of the integration over $%
\varepsilon$, only this term acquires the additional temperature damping
factor
\begin{equation}
R_{T}=(2\pi^{2}k_{B}T/(\hbar\omega_{c}))/\sinh\left(2%
\pi^{2}k_{B}T/(\hbar\omega_{c})\right),  \label{eq:RT}
\end{equation}
and the electron energy $\varepsilon$ is replaced by the chemical potential $%
\mu$. The macroscopic spatial inhomogeneities smear the Fermi energy along
the whole sample. Hence, in addition to the temperature smearing in Eq. (\ref%
{eq:sigmaT}), given by the integration over electron energy $\varepsilon$,
conductivity $\sigma$ acquires the coordinate smearing, given by the
integration over Fermi energy $\mu$ around its average value $\mu_0$ weighted by a normalized
distribution function $D\left(\mu \right) =D_0\left[\left(\mu -\mu_0 \right)
/W\right]$ of width $W$ : $\sigma=\int d\mu\,\sigma(\mu)D\left(\mu\right)$. 
Again, only the second term $\sigma_{xx}^{qo}$, describing MQO, is a rapidly
oscillating function of $\mu$ via $\overline{\alpha}\left(\mu\right)$, and only
this term acquires additional damping factor 
\begin{equation}
R_{W} =\int dx D_0\left( x\right) \cos \left( 2\pi x W/(\hbar\omega_{c})
\right) =R_{W}(W/(\hbar\omega_{c})),  \label{eq:RW}
\end{equation}
due to the sample inhomogeneities. This damping of MQO by long-range sample
inhomogeneities in layered organic metal $\beta$-(BEDT-TTF)$_{2}$IBr$_{2}$
was shown to be much stronger than the damping by usual short-range
impurities\cite{SO}, making the amplitude of SlO much larger than of MQO.
The SlO, given by Eq. (\ref{eq:sSlO}), do not depend on $\mu $ and,
hence, are not damped by the factor $R_W$. This property makes the
observation of SlO much easier than of MQO. It was used in the alternative
interpretation\cite{MQOYBCOJETPL,MQOYBCOPRB} of the observed\cite%
{HusseyNature2003,ProustNature2007,SebastianNature2008,AudouardPRL2009,SingletonPRL2010,SebastianPNAS2010,SebastianPRB2010,SebastianPRL2012,SebastianNature2014,ProustNatureComm2015}
MQO in YBCO high-temperature superconductors.

\subsection{Influence of electron spin on conductivity}

All previous expressions are for spinless electrons. If we take into account
the spin splitting of Fermi level $E_F\pm \frac{1}{2}g\mu_e|\boldsymbol{B}|$ ($g$
is the electron $g$-factor, $\mu_e$ is the Bohr magneton) and sum 
expressions (\ref{eq:sxxDrude}) for Drude conductivity over both spin
components, we simply multiply the spinless result (\ref{eq:sxxDrude}) by
two: 
\begin{equation}
\overline{\sigma}_{xx}^{(0)}(\varepsilon)\approx\frac{ e^{2}}{2\pi\hbar d}\frac{%
\overline{\alpha}\gamma_{0}}{\gamma_{0}^{2}+\pi^{2}}.
\end{equation}
For MQO $\sigma _{xx}^{qo}$, given by Eq. (\ref{eq:s1qo}), the sum of both
spin components gives 
\begin{gather}
\sigma _{xx}^{qo}(\varepsilon )\approx-2\overline{\sigma}_{xx}^{(0)}R_{D}R_S\times
\notag \\
\times\left[ \frac{2\pi ^{2}J_{0}\left( \lambda \right) }{\gamma
_{0}^{2}+\pi ^{2}}\cos \left( \overline{\alpha}\right) -\frac{ \lambda }{\overline{%
\alpha}}J_{1}(\lambda )\sin (\overline{\alpha})\right] ,  \label{eq:s1qoS}
\end{gather}
where the spin damping factor $R_S$ of MQO in quasi-2D metals with $t_z\ll
E_F$ is $R_{S}=\cos \left[ \pi gm_{\ast }/(2m_{e}\cos \theta) \right]$ ($m_e$
is the free electron mass).

The influence of spin splitting on SlO depends on electron dispersion and on
the coupling between two spin components. For the parabolic in-plane
dispersion, given by Eq. (\ref{eq:e3D0}), and in the absence of any coupling
between two spin components, the Zeeman spin splitting only adds a factor of 
$2$ to $\sigma_{xx}^{so}$, similar to the Drude term. Indeed, the SlO term
in Eq. (\ref{eq:sSlO}) does not depend on energy, and the sum over two
spin-split energy bands only adds a factor of $2$ to final expression.
However, for a more complicated in-plane electron dispersion this simple
conclusion may violate. Moreover, in real compounds there is often some
coupling between two spin components due to spin-dependent scattering,
chemical-potential oscillations and oscillating magnetostriction, or other
effects. This coupling between two spin components introduces additional
terms to SlO, which may lead to the angular dependence of SlO amplitude and
even to an analogue of the spin-zero effect.

\subsection{The limiting cases of large and small interlayer transfer integrals $t_z$}

In this section we compare the results obtained with two previously know
limiting cases, namely, 2D and 3D. The SlO are specific to quasi-2D metals,
being neglected in both these limiting case. In 2D case, $t_{z}=0$, the SlO
have zero frequency and, hence, do not exist. In 3D metals, where $t_{z}\sim
E_{F}\gg \hbar \omega _{c}$, the SlO may exist but have too small amplitude,
being less than MQO by a factor $\sim R_{D}\sqrt{\hbar \omega _{c}/(2\pi^2
t_{z})} \ll 1$. Hence, below we compare only the usual MQO of
intralayer conductivity.

In the 2D limiting case, taking $t_{z}=0$ and $\lambda =0$ in Eq. (\ref%
{eq:s1qo}), we obtain the following expression for the MQO of intralayer
conductivity 
\begin{equation}
\sigma _{xx}^{qo}(\varepsilon,\,t_{z}=0)\approx -\overline{\sigma}_{xx}^{(0)}R_{D}%
\frac{4\pi ^{2}}{\gamma _{0}^{2}+\pi ^{2}}\cos \left( \overline{\alpha}\right) .
\label{eq:be}
\end{equation}
It coincides with Eq. (2.15) of Ref. [\onlinecite{AndoIV}], where the quantum
transport in a 2D electron system under magnetic fields was studied. Note
that the amplitude of MQO in this Eq. (2.15) of Ref. [\onlinecite{AndoIV}] is twice
larger than in Eq. (2.16) of the same work\cite{AndoIV}
or in Eq. (6.40) of Ref. [\onlinecite{AndoFowlerSternRMP82}], where the quantum
oscillations of $\text{Im}\Sigma $ or $\tau $ are neglected.

The limiting 3D case corresponds to large $t_{z}\gg \hbar \omega _{c}$, i.e., 
$\lambda \gg 1$. In this limit one may use asymptotic expansions of the
Bessel functions at large argument in Eq. (\ref{eq:s1qo}): $J_0(\lambda)\approx\sqrt{2/(\pi\lambda)}\cos(\lambda-\pi/4)$, $J_1(\lambda)\approx\sqrt{2/(\pi\lambda)}\sin(\lambda-\pi/4)$. Then Eq. (\ref{eq:s1qo})
simplifies to 
\begin{gather}
\sigma _{xx}^{qo}(\varepsilon )\approx -\left( \frac{2^{3}}{\pi \lambda }%
\right) ^{1/2}\overline{\sigma}_{xx}^{(0)}R_{D}\left[ \frac{2\pi ^{2}\cos \left( 
\overline{\alpha}\right) }{\gamma _{0}^{2}+\pi ^{2}}\cos \left( \lambda -\frac{%
\pi }{4}\right) \right. -  \notag \\
\left. -\frac{\lambda }{\overline{\alpha}}\sin (\overline{\alpha})\sin \left( \lambda -%
\frac{\pi }{4}\right) \right] .  \label{eq:ba}
\end{gather}%
In a strong magnetic field $\gamma _{0}\ll 1$, $R_{D}\approx 1$, and Eq. (%
\ref{eq:ba}) in terms of initial parameters reduces to 
\begin{gather}
\frac{\sigma _{xx}^{qo}(\varepsilon )}{\overline{\sigma}_{xx}^{(0)}}\approx -%
\frac{2}{\pi }\left( \frac{2\hbar \omega _{c}}{t_{z}}\right) ^{1/2}\times  
\notag \\
\times \left[ \cos \left( \frac{2\pi E_{F}}{\hbar \omega _{c}}\right) \cos
\left( \frac{4\pi t_{z}}{\hbar \omega _{c}}-\frac{\pi }{4}\right) \right. - 
\notag \\
\left. -\frac{2t_{z}}{E_{F}}\sin \left( \frac{2\pi E_{F}}{\hbar \omega _{c}}%
\right) \sin \left( \frac{4\pi t_{z}}{\hbar \omega _{c}}-\frac{\pi }{4}%
\right) \right] .  \label{eq:meph}
\end{gather}%
We compare Eq. (\ref{eq:meph}) with the expression obtained in Ref. [\onlinecite{LifhitzKosevich}] (see Eq. (4) of Ref. [\onlinecite{LifhitzKosevich}]) and written in a more convenient form in Eq. (90.22) of the textbook\cite{LandauLifhitzKinetiks}: 
\begin{equation}
(\sigma _{xx})_{A}=\sum_{ex}\sum_{l=1}^{+\infty }(-1)^{l}\sigma
_{xx}^{(l)}\cos \left\{ l\frac{cS_{ex}}{e\hbar B_{z}}\pm \frac{\pi }{4}%
\right\} ,  \label{eq:sxxA}
\end{equation}%
where $ex\equiv\{min,\,max\}$ means extremal cross-section of the Fermi surface,
\begin{equation}
\sigma _{xx}^{(l)}=\frac{2^{5/2}\pi ^{1/2}(e\hbar )^{1/2}b_{ex}}{%
c^{1/2}B_{z}^{3/2}l^{1/2}}\left\vert \frac{\partial ^{2}S}{\hbar
^{2}\partial k_{z}^{2}}\right\vert _{ex}^{-1/2},  \label{eq:sl}
\end{equation}%
$b_{ex}$ is the quantity $b_{z}(E_{F},\,k_{z\,ex}(E_{F}))$ given by Eqs.
(90.13) and (90.15) of the book \cite{LandauLifhitzKinetiks} and taken at
points $k_{z\,ex}$, corresponding to Fermi surface extremal cross sections.
The ``$\pm $'' in Eq. (\ref{eq:sxxA}) means ``$-$'' for maximum and ``$+$'' for minimum of the function $S_{ex}(k_z)$\footnote{After Eq. (90.22) of the book \cite{LandauLifhitzKinetiks} it is mistakenly written ``$+$'' for maximum and ``$-$'' for minimum, which differs from the sign in the original paper \cite{LifhitzKosevich} and in our calculations.}.

In our case there are two extremal cross sections over the period $2\pi /d$.
These extremal cross section areas of the Fermi surface are 
\begin{equation}
S_{ex}=2\pi m_{\ast }(E_{F}+2t_{z}\cos (k_{z}d))|_{ex}=2\pi m_{\ast
}(E_{F}\pm 2t_{z}).
\end{equation}
Their second derivatives at extremal points are 
\begin{equation}
\frac{\partial ^{2}S}{\hbar ^{2}\partial k_{z}^{2}}|_{ex}=-\frac{4\pi
m_{\ast }d^{2}t_{z}}{\hbar ^{2}}\cos (k_{z}d)|_{ex}=\mp \frac{4\pi m_{\ast
}d^{2}t_{z}}{\hbar ^{2}}\label{eq:stalin}.
\end{equation}
If we assume that $b_{max}=b_{min}$, which is valid at least if $t_{z}\ll
E_{F}$, the sum over extremal cross sections for $l=1$ in Eq. (\ref{eq:sxxA}) can be
simplified: 
\begin{equation}
\sum_{ex}\cos \left\{ \frac{cS_{ex}}{e\hbar B_{z}}\pm \frac{\pi }{4}\right\}
=2\cos \left( \frac{2\pi E_{F}}{\hbar \omega _{c}}\right) \cos \left( \frac{%
4\pi t_{z}}{\hbar \omega _{c}}-\frac{\pi }{4}\right) .\label{eq:kim-jong-un}
\end{equation}
Using auxiliary Eqs. (\ref{eq:sl}), (\ref{eq:stalin}), and (\ref{eq:kim-jong-un}) in Eq. (\ref{eq:sxxA}), we find the oscillating part of intralayer conductivity for the first harmonic $l=1$
\begin{equation}
(\sigma _{xx}^{qo})_{A}\approx -\sqrt{\frac{2^5e\hbar^3b_{max}^2}{m_\ast c B^3_zt_zd^2}}\cos \left( \frac{2\pi
	E_{F}}{\hbar \omega _{c}}\right) \cos \left( \frac{4\pi t_{z}}{\hbar \omega
	_{c}}-\frac{\pi }{4}\right) .  \label{eq:LLQOA}
\end{equation}
From the Eq. (90.15) on p. 387 of the book\cite{LandauLifhitzKinetiks} one
can evaluate the intralayer conductivity $\sigma _{xx}$ averaged over the
period of magnetic oscillations 
\begin{equation}
(\overline{\sigma}_{xx})_{A}=\frac{2\hbar }{B_{z}^{2}}\int_{-\pi /d}^{\pi
/d}b(E_{F},\,k_{z})dk_{z}\approx \frac{4\pi \hbar b_{max}}{dB_{z}^{2}}.
\label{eq:sAv}
\end{equation}%
Finally, gathering Eqs. (\ref{eq:LLQOA}) and (\ref{eq:sAv}),
we find the ratio of oscillating and non-oscillating
parts: 
\begin{equation}
\frac{(\sigma _{xx}^{qo})_{A}}{(\overline{\sigma}_{xx})_{A}}\approx -
\left( \frac{2\hbar \omega _{c}}{\pi^2 t_{z}}\right) ^{1/2}\cos \left( \frac{2\pi
E_{F}}{\hbar \omega _{c}}\right) \cos \left( \frac{4\pi t_{z}}{\hbar \omega
_{c}}-\frac{\pi }{4}\right) .  \label{eq:LLQO}
\end{equation}%
This is twice smaller than in Eq. (\ref{eq:meph}), because in the derivation of Eqs. (\ref{eq:sxxA}--\ref{eq:LLQO}) the quantum oscillations of $b_{ex}$, and, hence, of $\text{Im}\Sigma $ are neglected. This extra factor of two, arising from the oscillations of $\text{Im}\Sigma $, is similar to that in the 2D case discussed above. If we neglected the MQO of $\text{Im}\Sigma $, instead of Eq. (\ref{eq:s1qo}) we would use Eq. (\ref{eq:s1b0}). Then, performing similar expansion as in the derivation of Eqs. (\ref{eq:ba}) and (\ref{eq:meph}), from Eq. (\ref{eq:s1b0}) we obtain Eq. (\ref{eq:LLQO}) in the lowest order in $t_z/E_F$. 

\section{Discussion}

The calculations of intralayer conductivity in the previous section shows
that $\sigma_{xx}(\mu)$ can be divided into three parts: 
\begin{equation}
\sigma_{xx}(\mu,\,T)\approx\overline{\sigma}_{xx}^{(0)}(\mu)+\sigma_{xx}^{qo}(%
\mu)R_{T}R_{W}+\sigma_{xx}^{so}(\mu),  \label{eq:S3}
\end{equation}
where $\overline{\sigma}_{xx}^{(0)}(\mu)$ represents the nonoscillating part of
conductivity, given by Eq. (\ref{eq:sxxDrude}), $\sigma_{xx}^{qo}(\mu)$
describes the MQO of intralayer conductivity, given by Eq. (\ref{eq:s1qo}), 
and $\sigma_{xx}^{so}(\mu)$ describes the slow oscillations of intralayer
conductivity, given by Eq. (\ref{eq:sSlO}). The second term, representing MQO,
acquires two damping factors $R_{T}$ and $R_{W}$ from temperature and
macroscopic sample inhomogeneities.

\begin{figure*}[t]
\includegraphics[width=0.9\textwidth,height=1\columnwidth]{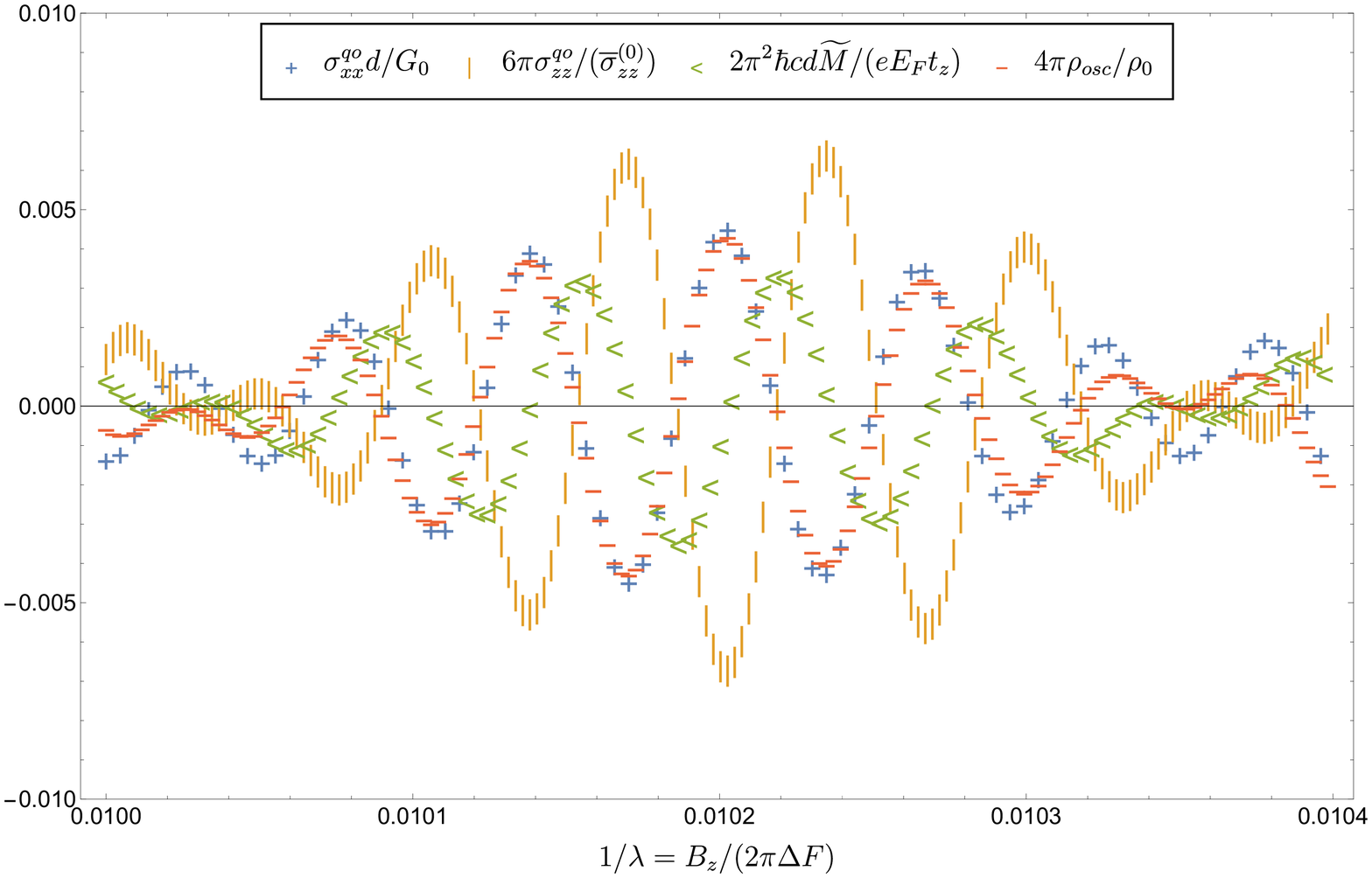}
\caption{Quantum oscillations of intralayer diagonal conductivity $\protect%
\sigma_{xx}^{qo}$, of interlayer conductivity $\protect\sigma_{zz}^{qo}$,
and of magnetization $\widetilde{M}$ as a
function $1/ \protect\lambda = B_{z}/(2\pi \Delta F)$. One can see that the oscillations
of $\protect\sigma_{zz}^{qo}$ are shifted from $
\widetilde{M}$ by a quarter of period, but at $1/\protect\lambda%
\approx0.01036$ the phase shift is close to $\protect\pi$. Except the beat nodes, the
oscillations of $\protect\sigma_{xx}$ have the same phase as those of the
density of states, but are in antiphase with the oscillations of $\protect%
\sigma_{zz}$. The parameters
for numerical calculations are: $m_{\ast}=0.04m_e$, $\Gamma _{0}=14.5\,K$, $%
t_{z}=10\,meV$, $E_{F}=200\,meV$.}
\label{fig:Quantum-oscillations-of}
\end{figure*}

The quantum oscillations of interlayer conductivity $\sigma_{zz}^{qo}(\mu )$, 
instead of Eq. (\ref{eq:s1qo}), 
are given by Eq. (18) of Ref. [\onlinecite{Shub}], which can be rewritten as 
\begin{equation}
\sigma_{zz}^{qo}(\mu)\approx2\overline{\sigma}_{zz}^{(0)}\cos\left(\overline{\alpha}%
\right)R_{D}\left[J_{0}\left(\lambda\right)-\frac{2}{\lambda}%
\left(1+\gamma_{0}\right)J_{1}\left(\lambda\right)\right].  \label{eq:szzqo}
\end{equation}
Let us compare Eq. (\ref{eq:s1qo}) for $\sigma_{xx}^{qo}$ with  Eq. (\ref{eq:szzqo}) for $\sigma_{zz}^{qo}$.
They look similar but have several important differences: (i) the total sign
``$-$'', responsible for the phase shift $\pi$ of MQO of in-plane $\sigma_{xx}$
with respect to interlayer $\sigma_{zz} $ conductivity, (ii) the amplitude
of MQO of $\sigma_{xx}$, given by Eq. (\ref{eq:rakshasa}), is nonzero even
in the beat nodes; (iii) additional field-dependent phase shift of MQO of
intralayer conductivity, given by Eq. (\ref{eq:dphi}), and (iv) the
expression in the square brackets in Eq. (\ref{eq:szzqo}), responsible for
the amplitude oscillations (beats) of MQO of interlayer conductivity,
contains extra term $\propto J_{1}\left(\lambda\right)$, which gives 
the field-dependent phase shift $\phi_{b}$ of beats of MQO
of $\sigma_{zz}$\cite{PhSh,Shub}. This phase shift $\phi_{b}$ contains the
parameter $2\left(1+\gamma_{0}\right)/\lambda=\left(1+\gamma_{0}\right)\hbar%
\omega_{c}/(2\pi t_{z}$), which is not small in strongly anisotropic Q2D
metals. This factor increases with the increase of magnetic field; it is $%
\sim 1$ in strongly anisotropic Q2D metals and $\ll 1$ in weakly anisotropic
almost 3D metals. For the in-plane conductivity $\sigma_{xx}$ in Eq. (\ref%
{eq:s1qo}) similar term results not in the phase shift of beats, but in the
phase shift of MQO themselves, given by Eq. (\ref{eq:dphi}). It is small by
the parameter $\lambda/\overline{\alpha}=2t_{z}/E_{F}$ and is approximately
field-independent. In strongly anisotropic Q2D metals $2t_{z}/E_{F}\ll1$,
and this phase shift is negligibly small. However, in weakly anisotropic Q2D
metals this parameter $\lambda/\overline{\alpha}=2t_{z}/E_{F}\sim1$, although
they have a cylindrical Fermi surface and are far from the Lifshitz
transition and magnetic breakdown, i.e., $E_{F}-2t_{z}\gg\hbar\omega_{c}$.

To measure the proposed phase shift of fast Shubnikov oscillations one can
compare the phase of Shubnikov and de Haas - van Alphen oscillations. The
latter are determined by the oscillations of DoS\cite{ChampelMineevPhilMag,Grig2001} 
\begin{equation}
\rho (B_{z})\approx \rho_0\left[ 1-2R_{D}J_{0}(\lambda )\cos \left( \overline{%
\alpha} \right) \right] \text{,}  \label{eq:DoS}
\end{equation}%
where the nonoscillating part of the DOS (per one spin) is $%
\rho_0=m_\ast/(2\pi\hbar^2d)$, and the magnetization oscillations per one spin component are given by\cite{Grig2001,Champel2001,Shub} 
\begin{gather}
\widetilde{{M}}(B_{z})\approx \frac{eE_F}{2\pi^2 \hbar c d}R_{D}R_{T} \times
\notag \\
\times\left( J_{0}(\lambda )\sin\left(\overline{\alpha}\right)+\frac{\lambda }{%
\overline{\alpha} }J_{1}(\lambda )\cos\left(\overline{\alpha}\right)\right) \text{.}
\label{eq:M}
\end{gather}%
Eqs. (\ref{eq:DoS}) and (\ref{eq:M}) are
illustrated in Fig. \ref{fig:Quantum-oscillations-of} and compared to
conductivity oscillations.

At low magnetic field, when $\lambda \gg 1$, the second term in the square
brackets of Eq. (\ref{eq:szzqo}) is small, and the MQO of $\sigma _{zz}$ in
Eq. (\ref{eq:szzqo}) and of $\sigma _{xx}$ in Eq. (\ref{eq:s1qo}) are in
antiphase. Note that the phase of $\sigma _{xx}$ MQO coincides with the
phase of DoS MQO given by Eq. (\ref{eq:DoS}). This is illustrated in Fig. %
\ref{fig:Quantum-oscillations-of}. However, at high fields ${\hbar \omega
_{c}\gg 4\pi t_{z}}$ expression (\ref{eq:szzqo}) for interlayer conductivity 
$\sigma _{zz}^{qo}$ asymptotically is equal to $-2\overline{\sigma}%
_{zz}^{(0)}\cos \left( \overline{\alpha}\right) R_{D}\left[ 1+2\gamma _{0}\right] 
$, while $\sigma _{xx}^{qo}$ is close to $-4\pi ^{2}\overline{\sigma}%
_{xx}^{(0)}R_{D}\cos \left( \overline{\alpha}\right) /\left( \gamma _{0}^{2}+\pi
^{2}\right) $. Hence, at high magnetic fields the fast oscillations of $%
\sigma _{xx}$, $\sigma _{zz}$ and DoS have the same phase. This agrees with
the calculations of $\sigma _{zz}$ within the two-layer model\cite%
{WIPRB2011,WIPRB2012,Grigbro,TarasPRB2014,TarasPRB2017} and for 3D
dispersion (\ref{eq:e3D}) at $\hbar \omega_{c}\gg t_{z}$\cite{ChampelMineev,GrigPRB2013}. Hence, there is a crossover between these two regimes of $\sigma _{zz}$ at $\lambda\sim 1$.

\begin{figure*}[tbh]
\subfloat[\label{fig:This-plot-demonstrates-1}Amplitude $A^{so}_{xx}$ of slow oscillations of normalized intralayer
	diagonal conductivity $\sigma_{xx}^{so}d/G_{0}$ as a function of $1/\gamma_0=\hbar\omega_c/(2\pi \Gamma_0)\propto B_z$. This plot demonstrates that the
	amplitude $A^{so}_{xx}$ changes its sign at $\gamma_{0}=\pi/\sqrt{3}$.]{\centering{}\includegraphics[width=0.85\textwidth,height=0.9\columnwidth]{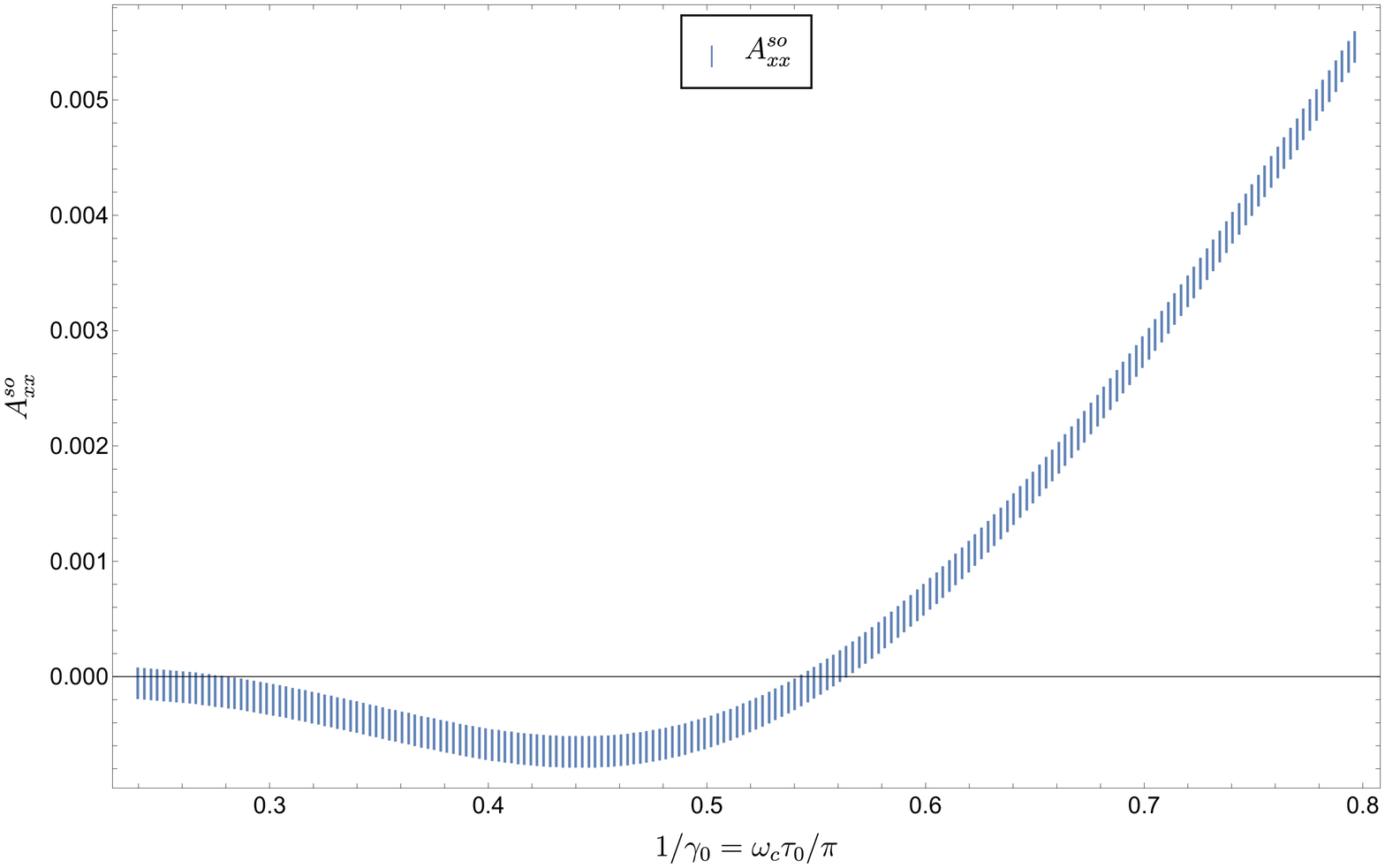}
	} \smallskip{} 
\subfloat[\label{fig:Slow-oscillations-of}Slow oscillations of intralayer $\sigma_{xx}$ and interlayer $\sigma_{zz}$ conductivity
	as a function of $1/\lambda =\hbar\omega_c/(4\pi t_z) 
	\propto B_{z}$. The slow oscillations of  $\sigma_{xx}$ and $\sigma_{zz}$ are in antiphase at low magnetic field and have the same phase at high field.]{ \centering{}\includegraphics[width=0.85\textwidth,height=0.9\columnwidth]{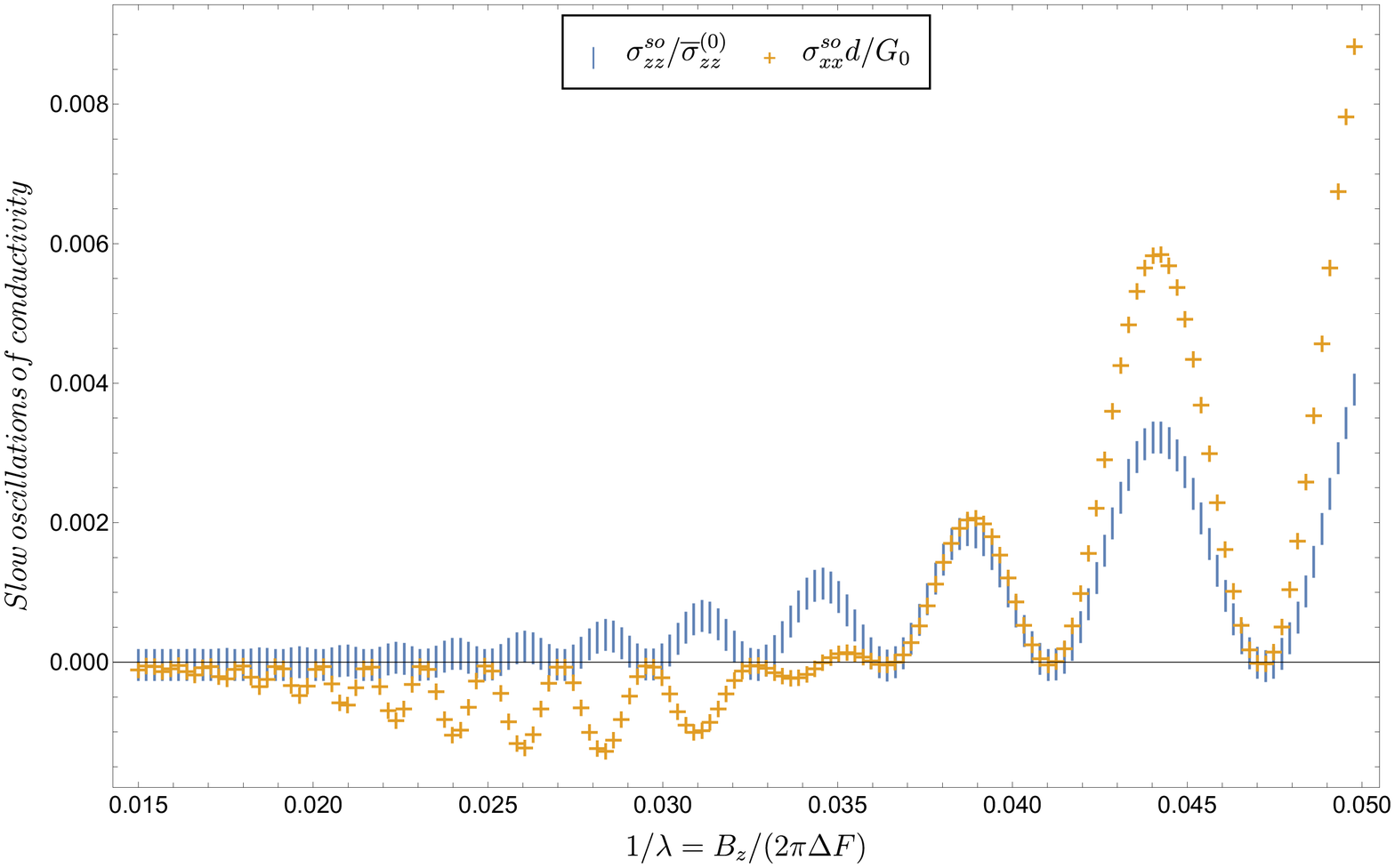}
	}
\caption{The amplitude (a) and magnitude (b) of slow oscillations of $%
\protect\sigma_{xx}$ and $\protect\sigma_{zz}$. The parameters are the same
as in Fig. \protect\ref{fig:Quantum-oscillations-of}.}
\label{fig:The-initial-data-1}
\end{figure*}

Let us now compare Eq. (\ref{eq:sSlO}) for $\sigma_{xx}^{so}(\varepsilon)$
with the slow oscillations of interlayer conductivity $\sigma_{zz}^{so}(%
\varepsilon)$, given by Eqs. (18) or (19) of Ref. [\onlinecite{Shub}], which can be
rewritten as 
\begin{equation}
\sigma_{zz}^{so}(\mu)\approx2\overline{\sigma}_{zz}^{(0)}R_{D}^{2}J_{0}\left(%
\lambda\right)\left[J_{0}\left(\lambda\right)-\frac{2}{\lambda}%
J_{1}\left(\lambda\right)\right].  \label{eq:szzSO}
\end{equation}
Similar to the beats of fast MQO, the slow oscillations of interlayer
conductivity $\sigma_{zz}$ have a field-dependent phase shift due to the
second term $2J_{1}\left(\lambda\right) /\lambda$ in the square brackets of
Eq. (\ref{eq:szzSO}), which is absent in the SlO of $\sigma_{xx}$. This
phase shift $\phi_{s}^{zz}\sim 2/\lambda $ is small at $\lambda \gg 1$, i.e.,
everywhere except the last period of slow oscillations.

The main difference of the SlO of interlayer $\sigma_{zz}^{so}$ and
intralayer $\sigma_{xx}^{so}$ conductivity is that the amplitude of the
latter depends nonmonotonically on $\gamma_{0}=\pi/(\omega_{c}\tau_0)$, as
one can see from Eq. (\ref{eq:sSlO}): at $\gamma_{0}^{2}=\pi^{2}/3$ the
amplitude of SlO of $\sigma_{xx}$ changes sign, going through zero. At small 
$\gamma_0 <\pi /\sqrt{3}$, i.e., at large $\omega_c\tau_0 >\sqrt{3}$ when MQO
are strong, the slow oscillations of intralayer $\sigma_{xx}$ and interlayer 
$\sigma_{zz}$ conductivity are in the same phase. At large $\gamma_0 >\pi /%
\sqrt{3}$, i.e., at small $\omega_c\tau_0 <\sqrt{3}$ when MQO are weak, the
SlO of $\sigma_{xx}$ and $\sigma_{zz}$ are in the antiphase. To demonstrate
this phase shift $\pi$ of SlO of in-plane conductivity $\sigma_{xx}$ with
respect to interlayer conductivity $\sigma_{zz}$, in Fig. \ref%
{fig:This-plot-demonstrates-1} we plot the amplitude 
\begin{equation}
A_{xx}^{so}\equiv\frac{\pi}{2\lambda}\overline{\alpha}\gamma_{0}R_{D}^{2}%
\frac{\pi^{2}-3\gamma_{0}^{2}}{(\gamma_{0}^{2}+\pi^{2})^{3}}  \label{eq:ASlO}
\end{equation}
of $\sigma_{xx}^{so}d/G_{0}$, where $G_{0}=e^{2}/(\pi\hbar)$ is the quantum
of conductance (the expression for the amplitude follows from the expression
of $\sigma_{xx}^{so}d/G_{0}$ after using the asymptote of squared Bessel
function $J^2_0(\lambda)\sim(1+\sin(2\lambda))/(\pi\lambda)$ for $%
\lambda\gg1 $ and extracting the coefficient before $\sin(2\lambda)$). In
Fig. \ref{fig:Slow-oscillations-of} we compare $\sigma_{xx}^{so}d/G_{0}$
given by Eq. (\ref{eq:sSlO}) to $\sigma_{zz}^{so}d/G_{0}$ given by Eq. (\ref%
{eq:szzSO}). Contrary to $\sigma_{xx}^{so}$, the amplitude of SlO of
interlayer conductivity $\sigma_{zz}^{so}$ in Eq. (\ref{eq:szzSO})
monotonically decreases with increasing $\gamma_{0}$ (see Fig. \ref%
{fig:The-initial-data-1}). The nonmonotonic field dependence of the
amplitude of slow oscillations of in-plane conductivity, probably, explains
the $\pi$-difference of the phase of SlO of in-plane magnetoresistance
observed\cite{GrigEuroPhys2016} in rare-earth tritellurides TbTe$_{3}$ and
GdTe$_{3}$ (see Fig. 6 of Ref. [\onlinecite{GrigEuroPhys2016}]).

\section{Summary}

To summarize, we calculate the magnetic quantum oscillations (MQO) of
intralayer conductivity $\sigma_{xx}$ in quasi-2D metals in quantizing
magnetic field. This calculation is based on the Kubo formula and harmonic
expansion. It takes into account the electron scattering by short-range
impurities and neglects the electron-electron interaction. The latter
approximation is justified in the metallic limit of large number of filled
LLs and finite interlayer transfer integral $t_z$. Previously, such
calculation in quasi-2D metals was performed only for interlayer
conductivity $\sigma_{zz}$\cite{ChampelMineev,Shub}. We calculated
analytically the amplitudes and phases of the usual MQO and the so-called
slow oscillations (SlO) with frequency $\propto t_z$, arising from the
mixing of two close MQO frequencies. The SlO appear only in the second order
in the Dingle factor, but they are usually stronger than MQO, because the
latter are additionally damped by temperature and sample inhomogeneities.

The comparison of the results for intralayer $\sigma_{xx}$ and interlayer $%
\sigma_{zz}$ conductivity shows several qualitative differences between
their oscillations, discussed and illustrated above. The amplitude
of SlO of $\sigma_{xx}$, given by Eqs. (\ref{eq:sSlO}) and (\ref{eq:ASlO}) and
illustrated in Fig. \ref{fig:The-initial-data-1}, 
has a nonmonotonic dependence on magnetic field. This amplitude changes sign at $%
\gamma_{0}=\pi/(\omega_{c}\tau_0) = \pi/\sqrt{3}$, while the amplitude of
SlO of $\sigma_{zz}$ is a monotonic function of field. The SlO of $%
\sigma_{xx}$ and $\sigma_{zz}$ have opposite phase in weak magnetic field
and same phase in strong field. The MQO of $\sigma_{zz}$ have a crossover with a phase
inversion at $\lambda\sim 1$, while MQO of $\sigma_{xx}$ do not have such
crossover. Therefore, similarly to SlO, the MQO of $\sigma_{zz}$ and $%
\sigma_{xx}$ have opposite phase in weak magnetic field and same phase in
strong field. This crossover between high- and low-field limits 
for MQO of $\sigma_{zz}$ is driven by the parameter 
$\lambda=4\pi t_{z}/(\hbar\omega_{c})$, while for SlO of $\sigma_{xx}$ the driving parameter
is $\gamma =2\pi \Gamma /(\hbar\omega_{c})$.

Notably, the oscillations of MQO amplitudes, called beats and arising from
the interference of two close frequencies, for $\sigma_{xx}$ are not
complete, i.e., the amplitude of $\sigma_{xx}$ oscillations is nonzero even
in the beat nodes, as given by Eq. (\ref{eq:rakshasa}) and illustrated in
Fig. \ref{fig:Amplitude-of-quantum}. The field-dependent phase shift of
beats, known for $\sigma_{zz}$ MQO\cite{PhSh,Shub}, does not appear in $%
\sigma_{xx}$. However, for $\sigma_{xx}$ the phase of MQO themselves is
shifted by the value $\,\sim t_z/E_F$, as given by Eqs. (\ref{eq:s1qo}), (%
\ref{eq:dphi}). 

The developed theory and the results obtained are applicable to describe
transverse magnetoresistance in various anisotropic quasi-2D conductors,
including organic metals, high-Tc superconducting materials,
heterostructures, intercalated graphite, rare-earth tritellurides, etc.

\acknowledgments{The authors thank the senior scientist Pavel
Streda from Department of Semiconductors of Institute of Physics of the
Czech Academy of Sciences for the detailed explanation of his calculations. 
T. I. M. thanks Konstantin Nesterov and Pavel Nagornykh for 
stylistic corrections.
T. I. M. acknowledges the RFBR grants \# 18-32-00205, 18-02-01022. 
P. G. acknowledges the program 0033-2018-0001 ``Condensed Matter Physics''.}

\end{document}